\documentclass[aps,prd,twocolumn,nofootinbib,preprintnumbers,superscriptaddresss,showkeys,showpacs,amsmath,amssymb,include graphics]{revtex4-1}
\usepackage{graphicx}
\usepackage{amsmath,amssymb}
\usepackage{amsfonts}
\usepackage{xspace} 
\usepackage[usenames]{color}
\usepackage{dcolumn}
\usepackage{bm}
\usepackage{mathrsfs}
\usepackage[colorlinks=true]{hyperref}
\usepackage[all]{hypcap} 
\usepackage[utf8]{inputenc} 
\usepackage{slashed}
\usepackage{multirow}
\usepackage{rotating}
\usepackage{nicematrix}
\usepackage{makecell}
\usepackage[bbgreekl]{mathbbol}
\usepackage{relsize}
\usepackage{slashed}
\usepackage{yfonts}
\DeclareSymbolFontAlphabet{\mathbb}{AMSb}
\DeclareSymbolFontAlphabet{\mathbbl}{bbold}
\usepackage{tikz-feynman}

\begin{document}

\title{Quarkyonic picture of isospin QCD}

\author{Oleksii Ivanytskyi$^{1}$} \email{oleksii.ivanytskyi@uwr.edu.pl} 

%\author[0000-0002-4947-8721]{Oleksii Ivanytskyi}

\affiliation{$^{1}$Incubator of Scientific Excellence---Centre for Simulations of Superdense Fluids, University of Wrocław, Max Born place 9, 50-204, Wrocław, Poland}

\date{\today}

%%%%%%%%%%%%%%%%%%%%%%%%%%%%%%%%%%%%%%%%%%%%%%%%%%%%%%%%%%%
\begin{abstract}

We suggest a new look onto the thermodynamics of cold isospin quantum chromodynamics (QCD) matter described within the quarkyonic framework, when confining forces are essential only for the momentum states close to the Fermi sphere and lead to formation of the quark-antiquark correlations with quantum numbers of pions.
Within this picture cold isospin QCD matter is constituted by the Bose-Einstein condensate of pions and free (anti)quarks.
We develop a simple confining model of pions existing as the lowest energy $P$-state of in-medium three-dimensional harmonic oscillator and use it to show that under the conditions of isospin QCD quarkyonic matter can onset at densities significantly smaller than the overlap density of pions. 
Using the experimentally established pion charge radius we demonstrate that the onset density can be even lower than two nuclear saturation densities.
To analyze the lattice QCD data we develop a field-theory inspired model of quark-antiquark-pion matter with vector-isovector repulsion among pions and (anti)quarks.
The quark substructure of pions, which is one of the key elements of the model, is accounted for within the one-loop no-sea approximation and leads to an effective medium dependence of the pion mass. 
The latter is shown to be crucial for providing asymptotic vanishing of the pion density and pion condensate, which reflects pion dissociation due to asymptotic freedom of QCD.
The proposed quarkyonic picture of isospin QCD agrees with the results of perturbative calculations on reaching the conformal limit of QCD and with the data of lattice simulations, which supports its viability and physical relevance.
\end{abstract}
%%%%%%%%%%%%%%%%%%%%%%%%%%%%%%%%%%%%%%%%%%%%%%%%%%%%%%%%%%%
\maketitle

%%%%%%%%%%%%%%%%%%%%%%%%%%%%%%%%%%%%%%%%%%%%%%%%%%%%%%%%%%%
\section{Introduction}
\label{intro}

The modern theory of strong interaction, quantum chromodynamics (QCD), provides a theoretical basis for studying properties and phase diagram of strongly interacting matter \cite{Wilczek:1999ym,Rischke:2003mt,Stephanov:2006zvm}.
While the QCD phase diagram is expected to have a rich structure, only two of its regions are, probably, realized in the Universe. 
The first of them corresponds to hot QCD matter with small densities of baryon charge and is accessible in the experiments on relativistic heavy ion collisions \cite{Iancu:2012xa,Busza:2018rrf,Shuryak:2014zxa} as well as in numerical simulations on discrete space-time lattices \cite{Bazavov:2017dus,Bazavov:2017dsy,HotQCD:2018pds,Ratti:2018ksb,Borsanyi:2018grb,Borsanyi:2020fev,Guenther:2020jwe,Borsanyi:2020fev,Borsanyi:2021sxv}.
The second region can be probed in the interiors of neutron stars (NS), when temperatures can be safely approximated by zero and baryon density is high \cite{Fukushima:2010bq,Kurkela:2009gj,Baym:2017whm}.
While the NS matter has recently gained a significant attention due to the impressive progress in the observational techniques performed in the last decade \cite{Antoniadis:2013pzd,Miller:2019cac,Riley:2021pdl,Riley:2019yda,NANOGrav:2019jur,LIGOScientific:2017vwq,LIGOScientific:2018hze,LIGOScientific:2018cki,Miller:2021qha,Fonseca:2021wxt}, it cannot be probed with the lattice QCD methods due to the sign problem \cite{PhysRevB.41.9301}.

This explains an intent to overcome the sign problem, which stimulated a splash of interest to simplified versions of QCD, where the sigh problem is absent and the lattice simulations can be performed.
The first of them is obtained when the number of kinds of the strong charge referred to as color is artificially reduced from three to two \cite{Bornyakov:2020kyz,Boz:2019enj,Iida:2020emi}.
Another interesting case, which is considered in this work, corresponds to the physically relevant three-color QCD at finite isospin but zero baryon densities (IQCD) \cite{Son:2000xc,deForcrand:2007uz,Brandt:2017oyy,Brandt:2021yhc,Brandt:2022hwy,Abbott:2023coj,Abbott:2024vhj,Brandt:2024dle}.
Remarkably, the universal character of strong interaction allows transferring the knowledge about superdense matter between the regimes of IQCD and baryon rich isospin symmetric QCD matter.
This allows constraining the equation of state (EoS) of symmetric matter using the EoS of IQCD \cite{Fujimoto:2023unl}. 

At zero temperature IQCD favors $u$ and anti-$d$ quarks as fundamental degrees of freedom at positive isospin chemical potentials and their antiparticles otherwise.
Strange (anti)quarks are important only at high temperatures. 
The possible hadron excitations of cold IQCD are determined by its quark content. 
At finite temperatures charged pions, are the most important hadronic degrees of freedom since they are the lightest isospin nonsinglet composite particles, which carry most of the isospin charge of the hadron phase of IQCD.
At sufficiently high isospin densities pions form the Bose-Einstein condensate (BEC), which has been indicated in the lattice QCD simulations \cite{Brandt:2021yhc,Brandt:2022hwy,Brandt:2024dle,Abbott:2023coj}.
At zero temperatures the hadron phase of IQCD is constituted by the pion BEC only.

The central role of pions in the low energy IQCD is also explained by the fact that they are the pseudoscalar Goldstone bosons of chiral symmetry breaking.
This is reflected in chiral perturbation theory ($\chi$PT) \cite{Son:1998uk,Son:2000xc,Splittorff:2000mm,Loewe:2002tw,Carignano:2016rvs,Andersen:2023ofv}.
It is applicable as long as the isospin chemical potential is small compared to the chiral symmetry breaking scale, which can be estimated as the difference of the masses of the $SU(2)_{L+R}$ chiral partners, pions and $\sigma$ mesons.
At higher isospin chemical potentials a more accurate description of the IQCD matter is provided by various versions of the linear $\sigma$-model \cite{Kapusta:1981aa,Haber:1981ts,Andersen:2006ys,Andersen:2008cb,Shu:2007na}.
However, this approach ignores that at a certain density hadrons, which are composite particles, must undergo a dissociation to unbound fermionic (anti)quarks just because of the Pauli principle that forbids a too dense occupation of the state space with bound fermions. 
As a result, (anti)quark excitations should appear in dense IQCD matter.

The Nambu-Jona-Lasinio (NJL) model fills this cavity and explicitly accounts for quark degrees of freedom \cite{Barducci:2004tt,Ebert:2005cs,Ebert:2005wr,He:2005sp,Lawley:2005ru,Andersen:2007qv,Andersen:2007qv}.
At the same time, the NJL model treats the IQCD matter in the mean-field approximation and identifies pion condensate with the isospin condensate of quarks.
The latter is only related to the Bardeen-Cooper-Schrieffer (BCS) type correlations in the deconfined phase and, strictly speaking, should vanish in the hadron phase of IQCD.
Furthermore, nonvanishing isospin condensate generates pairing among (anti)quarks.
In this case squared speed of sound of two-flavor chiral quark models, which include the NJL one, asymptotically approaches 1/5
\footnote{The squared speed of sound asymptotically approaches 1 if the vector repulsion is also taken into account.} \cite{Ivanytskyi:2022oxv}.
This value is in tension with the conformal limit of cold QCD \cite{Kurkela:2009gj,Gorda:2021kme,Gorda:2021znl}, while recovering it requires accounting for the nonlocal character of interactions among (anti)quarks \cite{Ivanytskyi:2024zip}.

The quark-meson (QM) model distinguishes between the pion condensate and isospin condensate of quarks already at the mean-field level \cite{Chiba:2023ftg,Kojo:2024sca,Andersen:2025ezj,Brandt:2025tkg}.
However, it treats mesons as fundamental degrees of freedom and, consequently, ignores their dissociation at high densities.
This conceptual aspect of the QM model comes into tension with deconfinement of quarks, which is unavoidable at high densities due to asymptotic freedom of QCD \cite{Gross:1973id,Politzer:1973fx}.
In addition, the QM model predicts the squared speed of sound to approach its conformal value from above \cite{Andersen:2025ezj}, which contradicts the results of perturbative QCD \cite{Kurkela:2009gj,Gorda:2021kme,Gorda:2021znl}.

Improving the description provided by the $\chi$PT, linear $\sigma$-, NJL and QM models requires a unified approach to quark-hadron matter.
The concept of quarkyonic matter was proposed as a physically intuitive framework for such approach \cite{McLerran:2007qj,Hidaka:2008yy}.
The main assumption of the quarkyonic picture of QCD that quarks remain bound to hadrons up to quark chemical potentials $\mu_q$ being parametrically large compared to the QCD energy scale $\Lambda$ can be illustrated by considering the limit of large number of colors $N_c$ when the QCD coupling vanishes as $g_{\rm QCD}\propto1/N_c$, the 't Hooft limit.
At zero temperature, considered in the present paper, Debye screening of confining interactions is generated by quark loops only so the leading order squared Debye mass scales as $g_{\rm QCD}^2N_c\mu_q^2$.
The confining interactions play no role when the Debye mass exceeds the QCD energy scale. i.e. when $\mu_q\gtrsim\sqrt{N_c}\Lambda$.
Below this chemical potential quarks from the vicinity of their Fermi sphere are bound to hadrons, while the momentum states resided deeply inside of the Fermi sea remain unbound since the Pauli blocking effects among them disable the confining scatterings \cite{McLerran:2007qj,Hidaka:2008yy}.
A phenomenological consequence of this picture is that quarks and hadrons can coexist at sufficiently high densities, which is also suggested within the scenario of quark-hadron continuity \cite{Schafer:1998ef,Baym:2017whm,Kojo:2021ugu,Fujimoto:2023mzy}.
Another important outcome of quarkyonic picture is stiffening of the EoS of baryon rich matter due to filling the low momentum states by quarks, which leads to populating the high momentum states with nucleons and, consequently, reduces the effect of their relatively high mass.
This feature of quarkyonic matter has been used to model the EoS of NSs
\cite{McLerran:2018hbz,Zhang:2020jmb,Koch:2022act,Cao:2022inx,Park:2021hqb,Gao:2024jlp,Folias:2024upz,Pang:2023dqj,Poberezhnyuk:2023rct,Duarte:2023cki}, consistent with the observational constraints on the maximum mass of these astrophysical objects \cite{Antoniadis:2013pzd,Fonseca:2021wxt}.

In this work we apply the quarkyonic picture for modeling cold IQCD matter, which, to the best of our knowledge, is the first attempt in this direction.
It is worth mentioning, in Ref. \cite{Cao:2016ats} the quarkyonic IQCD matter was discussed in the large $N_c$ and asymptotically free
limits without constructing the corresponding EoS and comparing it to the lattice QCD data.

The paper is organized as follows. In the next section we present the quarkyonic picture of isospin QCD.
In Sec. \ref{sec3} we consider a simple harmonic oscillator model of quarks confined in pions and use it to estimate parameters of the onset of quarkyonic matter in IQCD.
The quark substructure of pions and the corresponding medium dependence of the pion mass are considered in Sec. \ref{sec4}.
Section \ref{sec5} is devoted to developing a quarkyonic model of the isospion QCD matter and analyzing its thermodynamic properties.
The conclusions are given in Sec. \ref{concl}

%%%%%%%%%%%%%%%%%%%%%%%%%%%%%%%%%%%%%%%%%%%%%%%%%%%%%%%%%%%
\section{Quarkyonic picture}
\label{sec2}

While lacking a first principle justification, the quarkyonic picture of QCD is based on the assumption that confining forces among (anti)quarks are important only within a momentum shell near their Fermi surfaces \cite{McLerran:2007qj}.
These forces cause formation of hadrons, which are nothing but correlations of (anti)quarks from the mentioned shell that have the corresponding quantum numbers.
This picture resembles the phenomenon of Cooper instability in degenerate electron gas with an attractive interaction, which leads to the appearance of short living boson correlations, Cooper pairs, formed by the electrons residing in the momentum shell near the Fermi surface with the thickness determined by the pairing gap \cite{Cooper:1956zz}.
In the case of quarks and hadrons this thickness should be of order of the QCD energy scale $\Lambda$, which also separates perturbative and nonperturbative momentum states of (anti)quarks.  

Applying this quarkyonic picture to cold IQCD, we notice that its flavor content reduces to $N_f=2$ flavors, when pions are the lightest isospin nonsinglet hadrons and, thus, are most important.
In this work the isospin of (anti)quarks and pions is conventionally set to be one half and one, respectively.
Heavier hadrons carrying isospin, e.g., charged $\rho$ mesons and nucleons, are not explicitly accounted for.
Below it is argued that they should not appear in the hadron phase of cold IQCD matter since the corresponding threshold chemical potentials exceed the onset chemical potential of quarkyonic matter, where the contributions of the (anti)quark correlations with the quantum numbers of charged $\rho$ mesons and nucleons are subdominant.
For definiteness we also consider the case of positive isospin densities, when the IQCD matter is constituted by positively charged pions, $u$ and anti-$d$ quarks.
At negative isospin densities these particles should be replaced by their antipartners.
It is also clear that IQCD matter is symmetric with respect to exchange of $u$ and anti-$d$ quarks.
One of the manifestations of this symmetry is the coincidence of the maximum momenta carried by them.
Below these momenta are denoted as $k_I$.

\begin{figure}[t]
\label{fig1}
\includegraphics[width=\columnwidth]{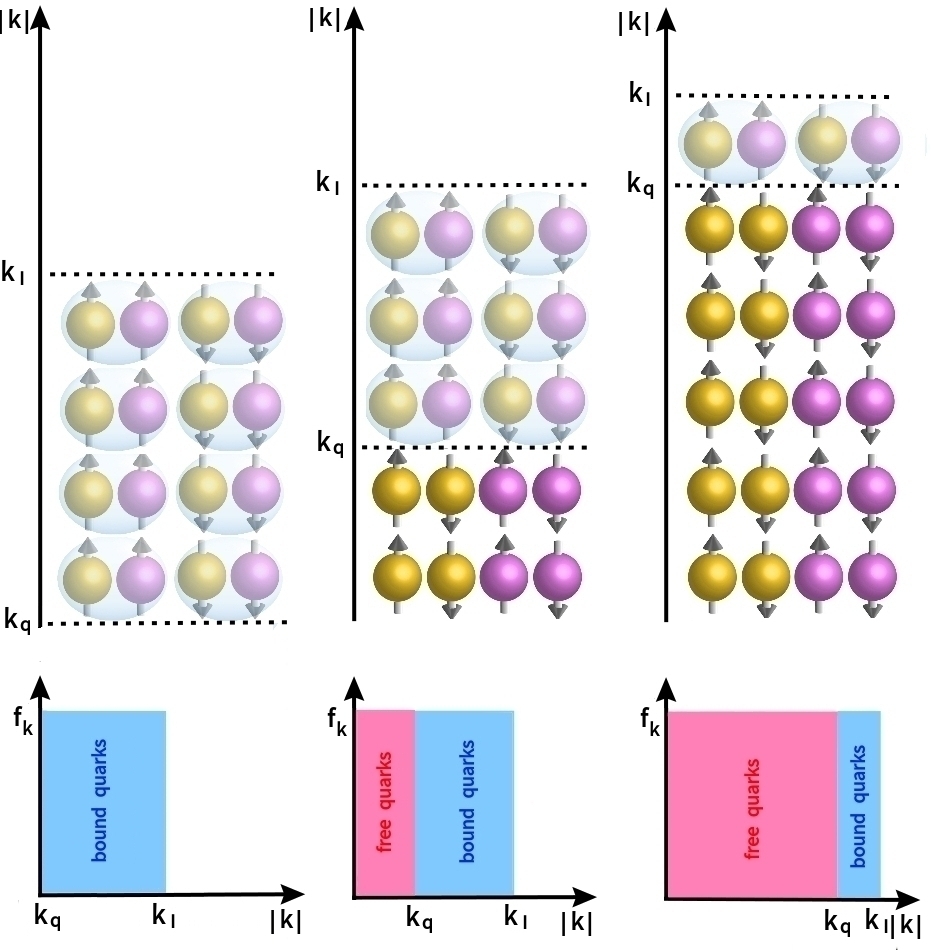}
\caption{Schematic representation of filling the quark momentum states at low isospin densities before the onset of quarkyonic matter (left panel), medium isospin densities above the onset of quarkyonic matter (middle panel) and high isospin densities deeply inside quarkyonic matter (right panel).
The golden and violet colors represent $u$ and anti-$d$ quarks of two possible spins indicated by the arrows.
The color states are not represented.
The bright spheres depict free quarks at momenta below $k_q$, while the dim spheres in blue transparent ellipses sketch the quark states with momenta between $k_q$ and $k_I$, which are bound to pions.
The blue and pink shaded areas in the sketches of the quark distribution function correspond to bound $f_{\bf k}^{\rm bound}$ and free $f_{\bf k}^{\rm free}$ quarks, respectively.}
\end{figure}

At small isospin densities, shown in the left panel of Fig. \ref{fig1}, $k_I<\Lambda/2$ so the maximum momentum transferred in quark scatterings does not exceed the QCD energy scale. 
As a result, all the (anti)quarks remain bound to pions.
Similarly to the Cooper pairs of electrons, pions appear as quark-antiquark correlations formed by scattering of the momentum states with the opposite three-momenta $\bf k$ and $\bf -k$.
Thus, pions have zero three-momentum.
At zero temperature, considered in this paper, this leads to formation of their BEC.

Increasing the isospin density leads to growth of the maximum momentum of (anti)quarks bound to pions, i.e., $k_I$.
At certain density the maximum momentum transferred in scatterings of these (anti)quarks $2k_I$ reaches the QCD energy scale $\Lambda$, which is enough to overcome the confining forces.
Above this density interaction among nonperturbative (anti)quark states gets too weak to confine them so they start dripping out of pions to the medium and form their own Fermi sphere by rescattering to low-momentum states.
The effects of Pauli blocking of these states prevent momentum exchange among them. 
As a result, these states omit confining interactions and remain free, as is illustrated in the middle panel of Fig. \ref{fig1}.
This corresponds to the quarkyonic picture of IQCD, when low-momentum (anti)quark states are free, while the high-momentum states are bound to pions.
Within this picture the onset density of quarkyonic picture $n_{\rm onset}$ is defined by the condition $k_I=\Lambda/2$.

Formation of the Fermi sphere of free (anti)quarks is also suggested within the picture of saturation of their distribution function \cite{Chiba:2023ftg,Kojo:2024sca}. 
According to it, the low-density distribution function of bound (anti)quarks has a Gaussian-like shape peaking at zero momentum with an amplitude below 1.
Increasing the isospin density leads to growth of this distribution function. 
The quarkyonic matter onsets when the occupation probability of the zero-momentum mode gets equal to 1.
Further increase of the isospin density extends the flat part of the distribution, which is followed by a smooth tail.
This flat part describes free (anti)quarks existing below the Fermi sphere of radius $k_q$.
This $k_q$ is nothing but the Fermi momentum of free quarks.

Above the onset density of quarkyonic matter the Fermi sphere of free (anti)quarks has a finite radius $k_q$, while the (anti)quarks bound to pions exist in the momentum shell between $k_q$ and $k_I$.
As is shown in the right panel of Fig. \ref{fig1}, at high isospin densities $k_q$ and $k_I$ get close so the momentum states bound to pions become less and less abundant.
A as result, the isospin density carried by pions decreases above the onset of quarkyonic matter.
This manifests dissociation of pions, which should asymptotically disappear due to asymptotic freedom of QCD \cite{Gross:1973id,Politzer:1973fx}.

In the general case, the momentum states below the quark Fermi sphere correspond to free $u$ and anti-$d$ quarks, while the states in the mentioned momentum shell are bound to pions.
The corresponding single-particle distribution functions are
\begin{eqnarray}
    \label{I}
    &&f_{\bf k}^{\rm free}=\theta\left(k_q-|{\bf k}|\right),\\
    \label{II}
    &&f_{\bf k}^{\rm bound}=
    \theta\left(|{\bf k}|-k_q\right)\theta\left(k_I-|{\bf k}|\right).
\end{eqnarray}
They are also sketched in Fig. \ref{fig1}.
The distribution functions of anti-$u$ and $d$ quarks vanish.

It is also worth mentioning, that the thickness of the pion shell is $k_I<\Lambda/2$ below the onset of quarkyonic matter, gets $k_I=\Lambda/2$ at this onset, attains a finite value $k_I-k_q<\Lambda/2$ above it and asymptotically vanishes at high densities. 
In Sec. \ref{sec5} this thickness is found in a self-consistent way.
Since only the momentum states from this shell are bound to pions, then only their dynamics determines pion properties.
In Sec. \ref{sec4} this circumstance is accounted for when considering pions as quark-antiquark correlations. 

%%%%%%%%%%%%%%%%%%%%%%%%%%%%%%%%%%%%%%%%%%%%%%%%%%%%%%%%%%%
\section{Onset of quarkyonic matter}
\label{sec3}

The quarkyonic matter onsets when the processes of quark exchange between pions become important and the quarks are no longer isolated from the medium.
In Refs. \cite{Chiba:2023ftg,Kojo:2024sca} the corresponding density $n_{\rm onset}$ was estimated to be about 5 nuclear saturation densities 
\footnote{In Refs. \cite{Chiba:2023ftg,Kojo:2024sca} the isospin of quarks and pions is one and two, respectively, which leads to $n_{\rm onset}$ about 10 nuclear saturation densities.}
, which is the overlap density of pions with the size defined by the charge radius $R_\pi=0.66~\rm fm$ \cite{PhysRevD.33.1785}. 

To show that quarkyonic matter can onset at smaller densities we consider a simple model representing pseudoscalar pions as the lowest energy state of a quark-antiquark system, which is bound by the three-dimensional harmonic potential $V=\mathfrak{m}_q\omega^2 r^2/2$ and provides the proper negative parity $\mathcal{P}_\pi=-1$.
Here $\mathfrak{m}_q$ is the reduced mass of (anti)quarks, $r$ stands for the spacial separation between them and $\omega$ defines the angular frequency of the oscillator.
Noticing that the pion parity can be expressed through the parities of quarks $\mathcal{P}_u=1$, antiquarks $\mathcal{P}_{\overline d}=-1$ and the orbital quantum number of the corresponding bound state $l$ as $\mathcal{P}_\pi=\mathcal{P}_u\mathcal{P}_{\overline d}(-1)^l$, we fix main, orbital and magnetic quantum numbers as $n=l=m=0$. 
This is equivalent to considering pion as the lowest $S$-state with
wave function normalized as $\langle\psi|\psi\rangle=1$ and the energy being 
\begin{eqnarray}
    \label{III}
    \psi=
    \left(\frac{\mathfrak{m}_q\omega}{\pi}\right)^{3/4}
    e^{-\frac{\mathfrak{m}_q\omega r^2}{2}}
    \quad{\rm and}\quad
    E=\frac{3\omega}{2}.
\end{eqnarray}
The oscillation frequency is adjusted so that mean separation between quarks $\langle\psi|r|\psi\rangle$ coincides with the charge radius of pions, i.e., we set $\mathfrak{m}_q\omega=4/\pi R_\pi^2$.

\begin{figure}[t]
\label{fig2}
\includegraphics[width=1\columnwidth]{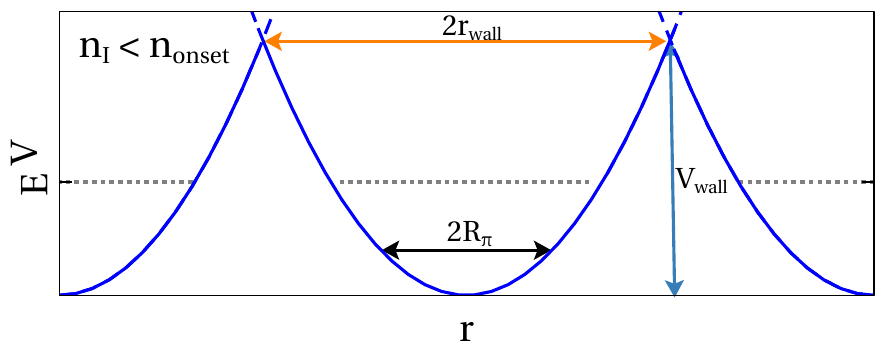}
\includegraphics[width=1\columnwidth]{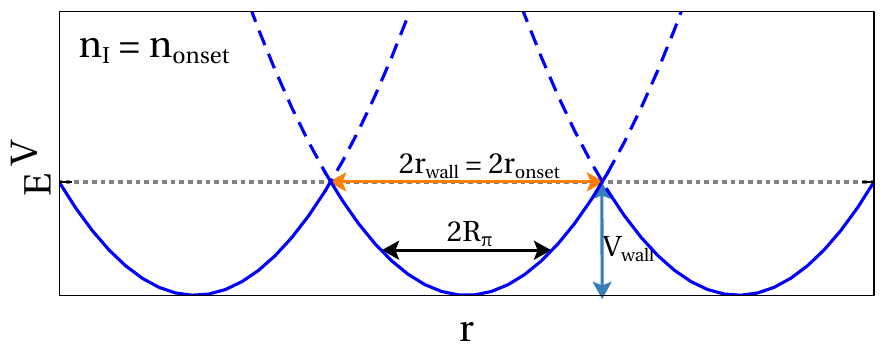}
\caption{Sketch of an ensemble of pions represented by the three-dimensional harmonic potentials below the onset of quarkyonic matter (upper panel) and at the onset (lower panel).
The solid and dashed branches of the curves represent the spacial domains, where the potentials do not and do overlap, respectively.
Intersections of the solid branches correspond to the tops of the potential walls of height $V_{\rm wall}$ separating (anti)quarks bound to different pions.
Twice the separation between the tops of the potential walls $2r_{\rm wall}$ is depicted by the orange horizontal arrows.
The horizontal dotted lines represent the pion energy $E$, which coincides with $V_{\rm wall}$ at the onset of quarkyonic matter when mean separation between quarks is $r_{\rm onset}=r_{\rm wall}$.
The black horizontal arrows show twice the charge radius of pions.}
\end{figure}

In medium harmonic potentials of pions partially overlap. 
This creates potential walls of a finite height $V_{\rm wall}$, which are separated by a mean distance $2r_{\rm wall}$ with $r_{\rm wall}=\sqrt{2V_{\rm wall}/\mathfrak{m}_q\omega^2}$ (see Fig. \ref{fig2} for details).
The height of the walls decreases with growth of density.
At small densities $V_{\rm wall}>E$ so (anti)quarks remain bound to pions and isolated from the medium.
At high densities $V_{\rm wall}<E$ and the processes of quark exchange between pions are possible.
The quarkyonic matter onsets when the height of the potential wall between pions $V_{\rm wall}$ coincides with the quark energy $E$.
In this case each pion occupies a spherical cell of the radius $r_{\rm onset}=r_{\rm wall}$ at $V_{\rm wall}=E$.
This $r_{\rm onset}=\sqrt{3\pi}R_\pi/2$ exceeds the pion charge radius.
This means that quarkyonic matter can onset before the overlapping of pions. 
The corresponding onset density is
\begin{eqnarray}
    \label{IV}
    n_{\rm onset}=\left(\frac{4}{3}\pi r_{\rm onset}^3\right)^{-1}=0.23~{\rm fm}^{-3}.
\end{eqnarray}
Note, accounting for the tunneling effects would reduce this estimate.
At the same time, the effects of pion repulsion and Pauli blocking can narrow the confining potential leading to higher values of the onset density.
We assume these competing effects to compensate each other so the estimate (\ref{IV}) remains valid.

With this $n_{\rm onset}$ we estimate the isospin chemical potential of the onset of quarkyonic matter $\mu_{\rm onset}$.
For this we use the leading order expression for the isospin density as a function of the isospin chemical potential obtained within the $\chi$PT \cite{Son:1998uk,Son:2000xc,Splittorff:2000mm,Loewe:2002tw,Carignano:2016rvs}, i.e., $n_I=f_\pi^2\mu_I(1-m_\pi^4/\mu_I^4)$, where $m_\pi=140$ MeV and $f_\pi=92$ MeV are mass and decay constant of pion in the vacuum fixed according to the Review of Particle Physics \cite{ParticleDataGroup:2022pth}. 
Solving this relation with $n_I=n_{\rm onset}$ as an equation with respect to $\mu_I=\mu_{\rm onset}$, we obtain
\begin{eqnarray}
    \label{V}
    \mu_{\rm onset}=238~{\rm MeV}.
\end{eqnarray}
This value immediately leads to the conclusion that such hadronic states as charged $\rho$ mesons and nucleons, which also carry isospin charge, do not appear in cold IQCD matter since their masses are higher than $\mu_{\rm onset}$.
Thus, the hadron phase of cold IQCD is constituted by the BEC of pions only.

According to the discussion above, the onset density of quarkyonic matter is nothing but a half of the number density of quarks bound to pions and existing under the Fermi sphere of radius $\Lambda/2$, i.e., $n_{\rm onset}=N_cN_f\Lambda^3/48\pi^2$.
Using this we find 
\begin{eqnarray}
    \label{VI}
    \Lambda=
    2\left(\frac{6\pi^2 n_{\rm onset}}{N_cN_f}\right)^{1/3}
    =519~{\rm MeV}.
\end{eqnarray}
This value agrees well with the values of the ultraviolet cutoff of chiral quark models of the NJL type \cite{Barducci:2004tt,Ebert:2005cs,Ebert:2005wr,He:2005sp,Lawley:2005ru,Andersen:2007qv,Andersen:2007qv,Scarpettini:2003fj,Blaschke:2007np,Hell:2008cc,Hell:2009by,Hell:2011ic,Ivanytskyi:2024zip} and is used below.

%%%%%%%%%%%%%%%%%%%%%%%%%%%%%%%%%%%%%%%%%%%%%%%%%%%%%%%%%%%
\section{Quark substructure of pions}
\label{sec4}

The quark substructure of pions can be accounted for by considering them as scattering states appearing via the quark-antiquark interaction with the quantum numbers of the isovector pseudoscalar particles.
This allows constructing the pion propagator as a four-point correlation function obtained after summing up a certain class of quark-antiquark scattering diagrams \cite{Klevansky:1992qe,Hatsuda:1994pi}. 
This propagator gives access to the spectral function of pions and their mass (see, e.g., Refs. \cite{Kitazawa:2013zya,Maslov:2023boq}).

As discussed in Sec. \ref{sec2}, only the quark states with $k_q<|{\bf k}|<k_I$ should be accounted for constructing the pion propagator.
This means that the Fermi momentum of free (anti)quarks $k_q$ plays the role of the infrared cutoff within quark loop diagrams.
Such infrared cutoff removes quark-antiquark thresholds from the integration interval of the quark loops even if pion mass exceeds two quark masses \cite{Blaschke:1998su}.
As a results, contrary to the chiral quark models of the NJL type \cite{Klevansky:1992qe,Hatsuda:1994pi,Kitazawa:2013zya,Maslov:2023boq}, within the quarkyonic picture the pion propagator has a pole even at high densities, when chiral symmetry gets restored.
This pole gives the access to the mass of stable pions $M_\pi$, which can be defined by the position of the pole evaluated at the rest frame.
Instead of thoroughly analyzing it within the quarkyonic picture, we aim at a simple physically motivated parametrization of the in-medium mass of pions, which effectively accounts for their quark substructure.
Analyzing the pole of the pion propagator at high isospin densities allows us to define the asymptote of $M_\pi$ analytically and parametrize the pion mass explicitly. 
Deriving this asymptote, we also adopt the no-sea approximation neglecting the quark zero point term, which is justified at high isospin densities.

We also account for the effects of running QCD coupling \cite{Deur:2016tte} by introducing the momentum dependent interaction vertex $i\gamma_5 g_{\bf k}$ motivated by the nonlocal chiral quark models in instantaneous separable approximation with three dimensional form factor $g_{\bf k}$ \cite{Scarpettini:2003fj,Blaschke:2007np,Hell:2008cc,Hell:2009by,Hell:2011ic,Ivanytskyi:2024zip}.
The latter can be chosen in various forms, which provide vanishing of $g_{\bf k}$ at high momenta \cite{Grigorian:2006qe}.
In the present work we use the Gaussian form of the form factor $g_{\bf k}=\exp(-{\bf k}^2/\Lambda^2)$.
As is shown later, this particular choice is not important for deriving the asymptote of the pion mass.

We start consideration with the leading order resummed propagator of pions obtained in the random phase approximation \cite{Klevansky:1992qe,Hatsuda:1994pi}
\begin{eqnarray}
    \label{VII}
    \mathcal{D}_\pi^{-1}=\frac{1}{2G_{\rm PS}}-\Pi_\pi.
\end{eqnarray}
The coupling $G_{\rm PS}$ controls strength of the interaction in the pseudoscalar channel and $\Pi_\pi$ is the polarization function of pions.
The propagator (\ref{VII}) can be also obtained within the Gaussian approximation to the chiral quark models \cite{Blaschke:2013zaa,Ivanytskyi:2022oxv}.

At the one-loop level one gets
\begin{eqnarray}
    \Pi_\pi&=&
    -\int\frac{d^4k}{(2\pi)^4}{\rm Tr}
    (i\gamma_5g_{\bf k}\mathcal{S}_k
    i\gamma_5g_{\bf p-k}\mathcal{S}_{p-k})\nonumber\\
    \label{VIII}
    \hspace*{-.1cm}&=&\hspace*{-.1cm}
    -8N_cN_f\int\frac{d{\bf k}}{(2\pi)^3}
    \frac{\epsilon_{\bf k} f_{\bf k}^{\rm bound}}{4\epsilon_{\bf k}^2-p^2}
    g_{\bf k}g_{\bf p-k},
\end{eqnarray}
where $\mathcal{S}_k$ is propagator of quarks with the single particle energy $\epsilon_{\bf k}=\sqrt{m_q^2+{\bf k}^2}$ and mass $m_q$, while $p$ is four-momentum of pions.
Unlike the chiral quark models, which define the medium-dependent quark mass within a self-consistent mean-field approximation, for the sake of simplicity we treat $m_q$ as a constant parameter of the model.
Note, in the general case $m_q$ does not coincide with the current quark mass from the Review of Particle Physics \cite{ParticleDataGroup:2022pth}.
Hereafter the trace is performed over the color, flavor and Dirac indexes.
The single particle distribution function $f_{\bf k}^{\rm bound}$ in Eq. (\ref{VIII}) ensures that only bound momentum states with  $k_q<|{\bf k}|\le k_I$ participate in forming pions in agreement with the quarkyonic picture described in Sec. \ref{sec2}.

It is worth mentioning that the polarization function given by Eq. (\ref{VIII}) does not include a contribution caused by the pion-pion interaction.
As is shown in Appendix \ref{secApp1} this contribution is negligible at small pion fractions, which is the case at high isospin densities.
This justifies ignoring the mentioned part of $\Pi_\pi$.

At high isospin densities quark masses in the expression for $\Pi_p$ can be neglected since $k_q\gg m_q$.
In addition, $k_I\rightarrow k_q$.
This allows us to explicitly perform integration and summation in Eq. (\ref{VIII}) and approximate the rest frame (${\bf p}=0$) polarization function of pions as
\begin{eqnarray}
    \label{IX}
    \Pi_p|_{k_q\rightarrow\infty}=-\frac{N_cN_fk_q(k_I-k_q)g_q^2}{\pi^2}
    \frac{4k_q^2}{4k_q^2-p^2},
\end{eqnarray}
where $g_q$ is the form factor defined at the Fermi momentum of free (anti)quarks.
Inserting this expression into Eq. (\ref{VII}), the asymptote of the pion mass is obtained as
\begin{eqnarray}
    \label{X}
    M_\pi^2|_{k_q\rightarrow\infty}=4k_q^2+\mathcal{O}(g_q^2).
\end{eqnarray}
As expected, the particular choice of the form factor does not affect this expression in the leading order.

Based on Eq. (\ref{X}) we propose the following parametrization of the effective pion mass
\begin{eqnarray}
\label{XI}  
    M_\pi^2=m_\pi^2+\frac{4k_q^4}{N_c^2\Lambda^2+k_q^2}.
\end{eqnarray}
The second term in Eq. (\ref{XI}) accounts for the quark substructure of pions by reproducing the high density asymptote of their mass $M_\pi^2\rightarrow4k_q^2$ discussed above.
The second term includes the additional factor $k_q^2/(N_c^2\Lambda^2+k_q^2)$ introduced in order to regularize the quark chemical potential in purely pion matter when $k_q=0$ (see Sec. \ref{sec5} for details).
In the absence of free quarks ($k_q=0$) the second term in Eq. (\ref{XI}) vanishes and pion mass attains its vacuum value. 

%%%%%%%%%%%%%%%%%%%%%%%%%%%%%%%%%%%%%%%%%%%%%%%%%%%%%%%%%%%
\section{Quarkyonic matter}
\label{sec5}

Within the picture described in Sec. \ref{sec2} number densities of pions and free (anti)quarks read
\begin{eqnarray}
\label{XII}
n_\pi&=&N_cN_f\int\frac{d{\bf k}}{(2\pi)^3}
f_{\bf k}^{\rm bound},\\
\label{XIII}
n_q&=&2N_cN_f\int\frac{d{\bf k}}{(2\pi)^3}f_{\bf k}^{\rm free}.
\end{eqnarray}
Note, the spin factor $2$ in the expression for $n_\pi$ is canceled since each pion consists of two (anti)quarks.
The total isospin density is
\begin{eqnarray}
\label{XIV}
n_I=n_\pi+\frac{n_q}{2}.
\end{eqnarray}
The factor $1/2$ in the term of (anti)quarks is nothing but their isospin.

For simplicity we consider the minimal scenario for pion matter composed of charged pions interacting via the exchange by neutral $\rho$ meson with constant mass $m_\rho=775$ MeV.
The corresponding coupling is denoted as $g$.
Such system is described by the Kroll-Lee-Zumino (KLZ) model \cite{Kroll:1967it}.
As argued in Sec. \ref{sec3}, this minimal scenario is justified by small chemical potential of the onset of quarkyonic matter.
At zero temperature pions, which carry the unit isospin charge, form the BEC.
The corresponding energy density reads
\begin{eqnarray}
    \label{XV}
    \varepsilon_\pi=
    n_\pi M_\pi+\frac{g^2n_\pi^2}{2m_\rho^2}.
\end{eqnarray}
The first term in this expression accounts for the single-particle contribution of condensed pions with zero momentum, while the second term is due to the $\rho$-meson mediated interaction treated at the mean-field level.
The pion condensate can be found as
\begin{eqnarray}
    \label{XVI}
    \langle\pi^*\pi\rangle=\frac{n_\pi}{2M_\pi}.
\end{eqnarray}
Details of derivation of $\varepsilon_\pi$ and $\langle\pi^*\pi\rangle$ are given in Appendix \ref{secApp1}.

The quark sector of the model is described within a quark model with vector-isovector repulsive interaction among (anti)quarks. 
A nonlocal treatment of this vector-isovector repulsion controlled by the coupling $G_V$ is required to provide asymptotically conformal behavior of quark matter \cite{Ivanytskyi:2024zip}.
The corresponding energy density is derived in Appendix \ref{secApp2} and reads
\begin{eqnarray}
\label{XVII}
\varepsilon_q=2N_cN_f\int\frac{d{\bf k}}{(2\pi)^3}
(\epsilon_{{\bf k}}+\phi g_{\bf k})f_{\bf k}^{\rm free}-
\frac{\phi^2}{4G_V}.
\end{eqnarray}
Here $\phi$ denotes the zeroth component of the auxiliary vector-isovector field.
The amplitude of this field is defined by minimizing the energy density of quarks, i.e.,
\begin{eqnarray}
\label{XVIII}
\phi=4G_VN_cN_f\int\frac{d{\bf k}}{(2\pi)^3}
g_{\bf k}f_{\bf k}^{\rm free}.
\end{eqnarray}
At hight densities $k_q\gg\Lambda$ and $g_{\bf k}f_{\bf k}^{\rm free}\simeq g_{\bf k}$. 
With this we obtain the high density asymptote of the vector-isovector field $\phi_\infty=2G_VN_cN_f\Lambda^3/\sqrt{\pi}$, which is finite.
This finite vector-isovector field is negligible compared to the Fermi momentum of free quarks as well as their mass.
Thus, the high density asymptote of the considered model of quark matter corresponds to the gas of free massless quarks, which respects the conformal limit of QCD.

The total energy density is a sum of the contributions of pions and (anti)quarks
\begin{eqnarray}
\label{XIX}
\varepsilon=\varepsilon_\pi+\varepsilon_q.
\end{eqnarray}

Chemical potentials of pions and (anti)quarks can be found as the derivatives of the total energy density with respect to the corresponding particle number densities, evaluated at constant number density of another component.
Since constant number density of free (anti)quarks requires a constant value of their Fermi momentum, then $\mu_\pi$ becomes
\begin{eqnarray}
    \label{XX}
    \mu_\pi=\frac{\partial\varepsilon}{\partial n_\pi}=M_\pi+\frac{g^2n_\pi}{m_\rho^2},
\end{eqnarray}
The first term in this expression represents single particle energy of pions in the BEC, while the seconds term accounts for the contribution generated by the $\rho$-meson mediated interaction among them. 
Given that the isospin of pions is one, we define the isospin chemical potential as $\mu_I=\mu_\pi$.

Equation (\ref{XX}) allows us to fix the coupling $g$ if the isospin density and isospin chemical potential at the onset of quarkyonic matter are known.
This yields
\begin{eqnarray}
   \label{XXI}
   \frac{m_\rho^2}{g^2}=\frac{n_{\rm onset}}{\mu_{\rm onset}-m_\pi},
\end{eqnarray}
which is the relevant parameter of the developed model.
Here we accounted for the fact that $M_\pi=m_\pi$ at the onset of quarkyonic matter.

Similarly, we find the quark chemical potential 
\begin{eqnarray}
    \label{XXII}
    \mu_q=\frac{\partial\varepsilon}{\partial n_q}
    =\epsilon_q+\phi g_q+n_\pi
    \frac{\partial M_\pi}{\partial n_q}.
\end{eqnarray}
The first two terms in Eq. (\ref{XXII}) are the Fermi energy of free (anti)quarks, which includes their mass and kinetic energy $\epsilon_q$ and the contribution caused by the repulsive interaction among them $\phi g_q$.
Positiveness of the derivative $\partial M_\pi/\partial n_q$ shows that quark chemical potential exceed their Fermi energy since adding a free (anti)quark on the top of their Fermi sphere shifts the pion shell and, consequently, changes the effective mass of pions. 
The third term in Eq. (\ref{XXII}) accounts for this effect.
It is clear that shifting the pion shell does not happen in the absence of quarks, i.e. at the onset of quarkyonic matter and below it. 
Therefore, we require $\partial M_\pi/\partial n_q=0$ at $k_q=0$.
This explains introducing the factor $k_q^2/(N_c^2\Lambda^2+k_q^2)$ in the second term in the parametrization of the effective pion mass (\ref{XI}).
This also provides the convergence of the quark chemical potential to $\mu_q=m_q$ in pure BEC of pions ($k_q=0$).
Otherwise, $\mu_q$ would diverge as $1/k_q$ in the absence of quarks.

The condition of equilibrium with respect to strong decays of pions to their constituent (anti)quarks can be written as
\begin{eqnarray}
    \label{XXIII}
    \mu_\pi=2\mu_q.
\end{eqnarray}
This condition insures that $\mu_q=\mu_I/2$ in agreement with the isospin of quarks being one half.
It also relates the Fermi momentum of free (anti)quarks $k_q$ to the maximum momentum of bound (anti)quarks $k_I$.
Treating $k_I$ as the parameter controlling the thermodynamic state of the IQCD matter and determining $k_q$ as described above, we can construct all the thermodynamic quantities of the quarkyonic IQCD matter.
Particularly, we define the total pressure through the thermodynamic identity $P=\mu_I n_I-\varepsilon$.
With this we arrive at speed of sound $c_s^2=\partial P/\partial\varepsilon=\partial\ln\mu_I/\partial\ln n_I$ and dimensionless interaction measure $\delta=1/3-P/\varepsilon$.

As a guidance, below we also consider pressure of the $\chi$PT $P=f_\pi^2(\mu_I^2-m_\pi^2)^2/2\mu_I^2$, which is recovered from the isospin density by integrating the thermodynamic identity $n_I=\partial P/\partial\mu_I$ and requiring $P=0$ at $\mu_I=m_\pi$, its energy density $\varepsilon=f_\pi^2(\mu_I^2+3m_\pi^2)(\mu_I^2-m_\pi^2)/2\mu_I^2$, pion condensate $\langle\pi^*\pi\rangle=n_\pi/2m_\pi$, dimensionless interaction measure $\delta=2(3m_\pi^2-\mu_I^2)/3(3m_\pi^2+\mu_I^2)$ and speed of sound $c_S^2=(\mu_I^4-m_\pi^4)/(\mu_I^4+3m_\pi^4)$.

At the quark onset when $k_q=0$ and $k_I=\Lambda/2$ the chemical potentials of quarks $\mu_q=m_q$ and pions $\mu_\pi=\mu_{\rm onset}$.
With this Eq. (\ref{XXIII}) allows us to exclude the quark mass from the list of the independent model parameters, i.e.
\begin{eqnarray}
    \label{XXIV}
    m_q=\frac{\mu_{\rm onset}}{2}.
\end{eqnarray}
Above the quarkyonic matter onset ($k_I>\Lambda$) quark Fermi momentum $k_q$ can be found using Eq. (\ref{XXIII}). 
It is clear that $k_q=0$ at $k_I\le\Lambda/2$.

\begin{table}[t]
\centering
\begin{tabular}{|c|c|c|c|c|c|c|c|c|}
\hline    
$m_\pi$ & $f_\pi$ & $R_\pi$ & $r_{\rm onset}$ &    $n_{\rm onset}$  &  $\mu_{\rm onset}$ &   $\Lambda$   & $m_\rho/g$  &    $m_q$      \\
 $[\rm MeV]$ & $[\rm MeV]$ & $[\rm fm]$  &   $[\rm fm]$   & $[\rm fm^{-3}]$  &    $[\rm MeV]$     &  $[\rm MeV]$  & $[\rm MeV]$ &   $[\rm MeV]$ \\ \hline
 140 & 92 & 0.66 &    1.011      &        0.231        &         238        &      519      &    135      &      119      \\ \hline
\end{tabular}
\caption{Values of the physical characteristics of pions, onset of quarkyonic matter and model parameters used in this work.} 
\label{table1}
\end{table}

Table \ref{table1} summarizes the parameters of the present model of IQCD.
We also consider several values of the vector-isovector coupling $G_V$, which is the only unfixed parameter of the model.
We confront the results of the modeling to the lattice data on thermodynamics of cold IQCD matter \cite{Abbott:2023coj}.

Before going further, we want to demonstrate that within the present model the pion fraction vanishes at high isospin density. 
For this we use Eqs. (\ref{XX}), (\ref{XXII}) and (\ref{XXIII}) to express $n_\pi$ and notice that at high densities $\epsilon_q\simeq k_q+m_q^2/2k_q$, $M_\pi\simeq2k_q+(m_\pi^2-4N_c^2\Lambda^2)/4k_q$.
With this we obtain
\begin{eqnarray} 
    \label{XXV}
    n_\pi=\frac{2(\epsilon_q+\phi g_q)-M_\pi}{\frac{g^2}{m_\rho^2}-\frac{\partial M_\pi}{\partial n_q}}
    \rightarrow
    \frac{m_\rho^2}{g^2}\frac{4m_q^2+4N_c^2\Lambda^2-m_\pi^2}{4k_q},
\end{eqnarray}
where the $n_I\rightarrow\infty$ limit is taken on the second step and the contributions of the vector-isovector field and the derivative $\partial M_\pi/\partial n_q$ were neglected since they are suppressed exponentially and $\propto1/k_q^2$, respectively. 
Equation (\ref{XXV}) shows that at high isospin density $n_\pi$ vanishes and the total isospin density can be approximated as $n_I\simeq n_q/2$.
Using the explicit expressions for $n_\pi$ and $n_q$ given by Eqs. (\ref{XII}) and (\ref{XIII}) we also find that at high isospin densities
\begin{eqnarray}
    \label{XXVI}
    k_q\overset{n_I\rightarrow\infty}{\longrightarrow}k_I-\frac{m_\rho^2}{g^2}\frac{4m_q^2+4N_c^2\Lambda^2-m_\pi^2}{12n_I}.
\end{eqnarray}
It follows from this relation that thickness of the pion shell in the momentum space vanishes $\propto1/n_I\propto 1/k_I^3$. 
This behavior agrees with the conclusion following from Eq. (\ref{XXV}) that pion density vanishes at high densities as $n_\pi\propto 1/k_I\simeq1/k_q\simeq2/\mu_I$.
Using this results and $M_\pi\simeq2k_q\simeq\mu_I$, we conclude that $\langle\pi^*\pi\rangle\propto1/\mu_I^2$ at high densities.

To find the high density asymptote of speed of sound we use the leading order expansion of the isospin chemical potential $\mu_I\simeq2k_q+m_q^2/k_q$ and isospin density $n_I\simeq n_q/2$.
Following Ref. \cite{Ivanytskyi:2024zip}, we treat the dimensionless interaction measure at high densities using the  L’H\^opital’s rule, which leads to $\delta\simeq1/3-c_S^2$.
This allows us to show that at high densities
\begin{eqnarray}
    \label{XXVII}
    c_S^2\simeq\frac{1}{3}\left[1-\frac{4m_q^2}{\mu_I^2}\right]
    \quad{\rm and}\quad
    \delta\simeq\frac{4m_q^2}{3\mu_I^2}.
\end{eqnarray}
Noticing that $\mu_q=\mu_I/2$, we conclude that these expressions coincide with the ones obtained for asymptotically conformal CFL quark matter with vector repulsion \cite{Ivanytskyi:2024zip}.
Thus, the proposed quarkyonic picture of IQCD predicts reaching the conformal limit $c_S^2=1/3$ and $\delta=0$ at asymptotically high densities.
Equation (\ref{XXVII}) demonstrates that similarly to the case of the CFL quark matter at high baryonic densities \cite{Ivanytskyi:2024zip} this limit is reached from below for $c_S^2$ and from above for $\delta$.
Such behavior is also predicted by perturbative QCD at zero temperature but high baryon densities \cite{Kurkela:2009gj,Gorda:2021kme,Gorda:2021znl}, while the QM model predicts the opposite behavior \cite{Chiba:2023ftg,Kojo:2024sca,Andersen:2025ezj,Brandt:2025tkg}.
In Ref. \cite{Chiba:2023ftg} this drawback of the QM model has been attributed to the lack of gluon exchanges and nonperturbative power corrections arising from the meson condensates.
The latter reflects a conceptual problem of the QM model, which treats mesons as fundamental degrees of freedom and, consequently, ignores their dissociation at high densities. 
This problem is absent within the quarkyonic picture of IQCD proposed in this work.

Analyzing the high density behavior of the present model also allows showing the importance of accounting for quark substructure of pions introduced to the present model via the mass scaling (\ref{XI}).
To demonstrate this we consider a fictitious situation when at high densities the effective pion mass grows slower than $\propto k_q$.
In this case $M_\pi$ and its derivative in Eq. (\ref{XXVI}) can be neglected, which leads to the linearly diverging asymptote of pion density $n_\pi\rightarrow m_\rho^2\mu_I/g^2$ and a diverging pion condensate.
The pion density also linearly diverges when at high densities the effective pion mass behaves as $2\zeta k_q$ with $\zeta<1$. Similarly to the QM model with the one-loop quark dressing of meson propagators, in this case the pion condensate saturates to the constant value $\langle \pi^*\pi\rangle\rightarrow(\zeta^{-1}-1)m_\rho^2/g^2$.
None of this possibilities is consistent with quark deconfinement via dissociation of pions when their density asymptotically vanishes.
The mass scaling (\ref{XI}) accounts for such dissociation within the quarkyonic picture of IQCD.

Repeating the analysis described above, we find that in the fictitious case of $M_\pi/k_q\rightarrow0$ the asymptote of speed of sound (\ref{XXVII}) receives the positive correction $16\pi^2 m_\rho^2/3N_cN_fg^2\mu_I^2$.
Using the values of the model parameters from Table \ref{table1} we conclude that in this case $c_S^2$ and $\delta$ reach their conformal values from above and below, respectively, which contradicts to the predictions of the perturbative QCD.
Thus, the mass scaling (\ref{XI}), which accounts for the quark substructure of pions, is crucial for providing such agreement.
This explains why the models ignoring this substructure predict $c_S^2\rightarrow1/3+0$ and $\delta\rightarrow-0$.

\begin{figure}[t]
\label{fig3}
\includegraphics[width=.9\columnwidth]{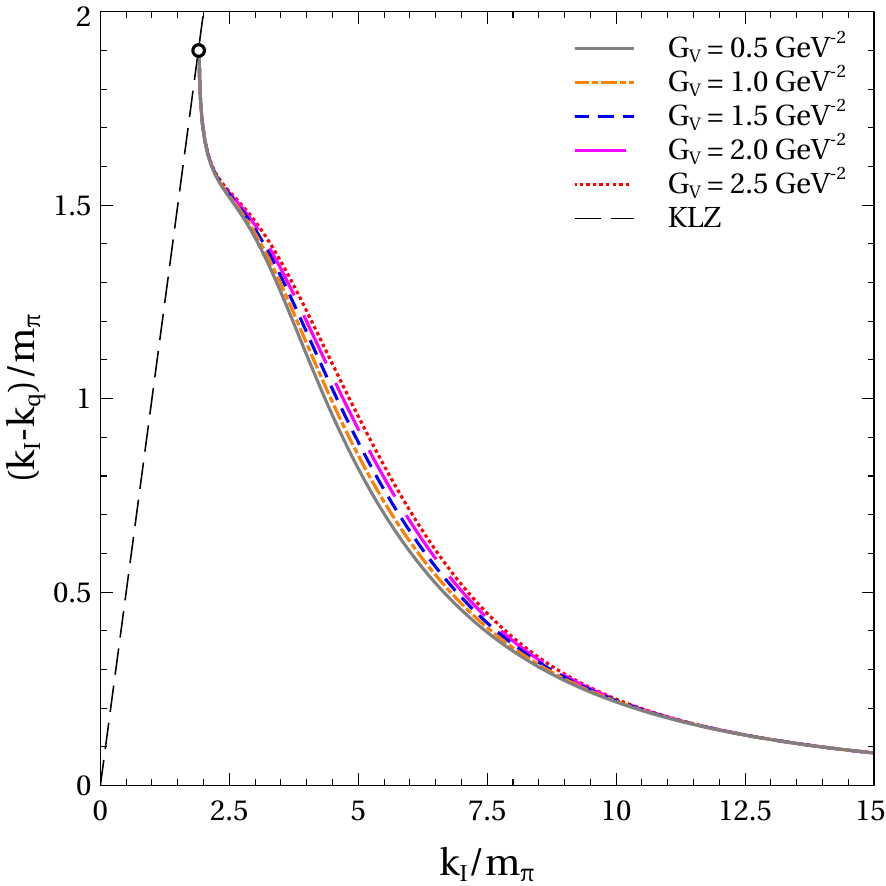}
\caption{Thickness of the pion shell $k_I-k_q$ as a function of the maximum momentum of bound quarks $k_I$ in units of the vacuum pion mass $m_\pi$.
The calculations are performed within the quarkyonic picture of IQCD with several values of the vector-isovector coupling $G_V$, indicated in the legend, and the KLZ model of interacting pion matter.
Empty circle indicates the onset of quarkyonic matter separating these two regimes.}
\end{figure}

We first consider the thickness of the pion shell as a function of the maximum momentum of bound quarks shown on Fig. \ref{fig3}. 
Below the onset of quarkyonic matter this thickness coincides with $k_I$, which reflects absence of free quarks and vanishing of their Fermi momentum $k_q$.
Above this onset $k_I-k_q$ monotonously decreases with the asymptotic behavior given by Eq. (\ref{XXVI}).
At the same time, above the onset of quarkyonic matter $k_I-k_q$  exhibits two inflection points and a shoulderlike structure at $k_I/m_\pi\simeq2.5-5$. 
Thus, behavior of the pion shell thickness is more complicated than just a power low dependence on the maximum momentum of bound quarks, which is commonly used at modeling cold baryon rich matter within the quarkyonic picture (see, e.g., Refs.  \cite{McLerran:2018hbz,Zhao:2020dvu,Gao:2024jlp}).
It is seen from Fig. \ref{fig2} that increasing the vector-isovector coupling leads to thicker pion shell.
This effect reflects the fact that stronger repulsion among free (anti)quarks increases the energy cost of their excitations.
As a result, the fraction of (anti)quarks bound to pions increases, which makes their shell thicker.

\begin{figure}[t]
\label{fig4}
\includegraphics[width=0.9\columnwidth]{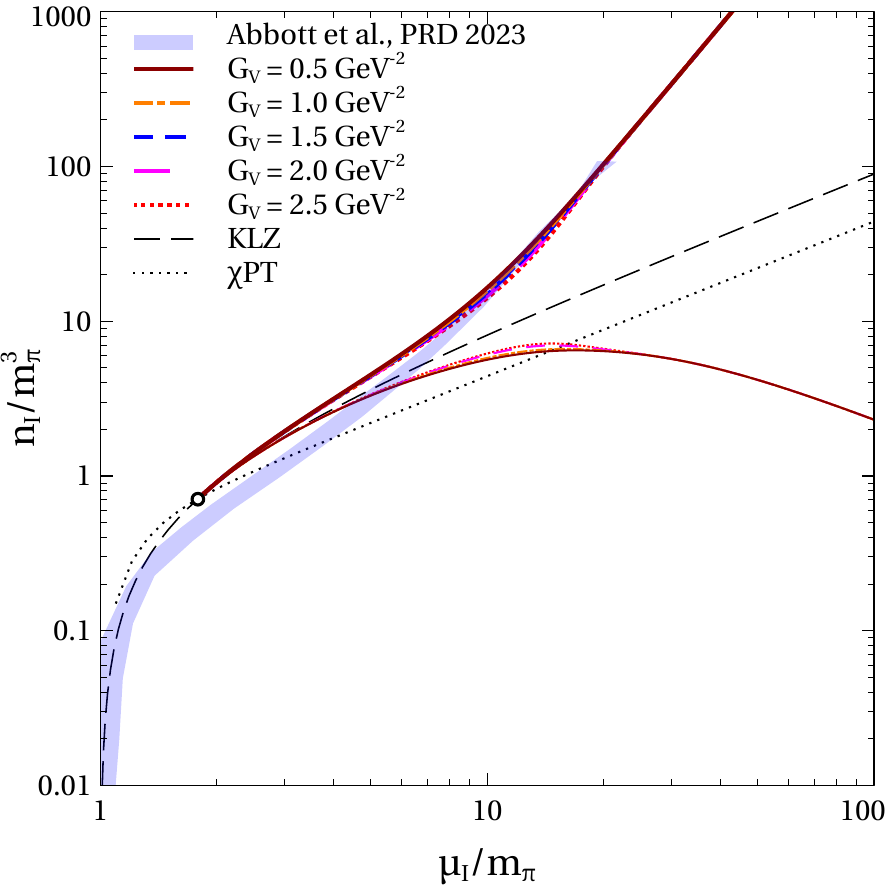}
\includegraphics[width=0.9\columnwidth]{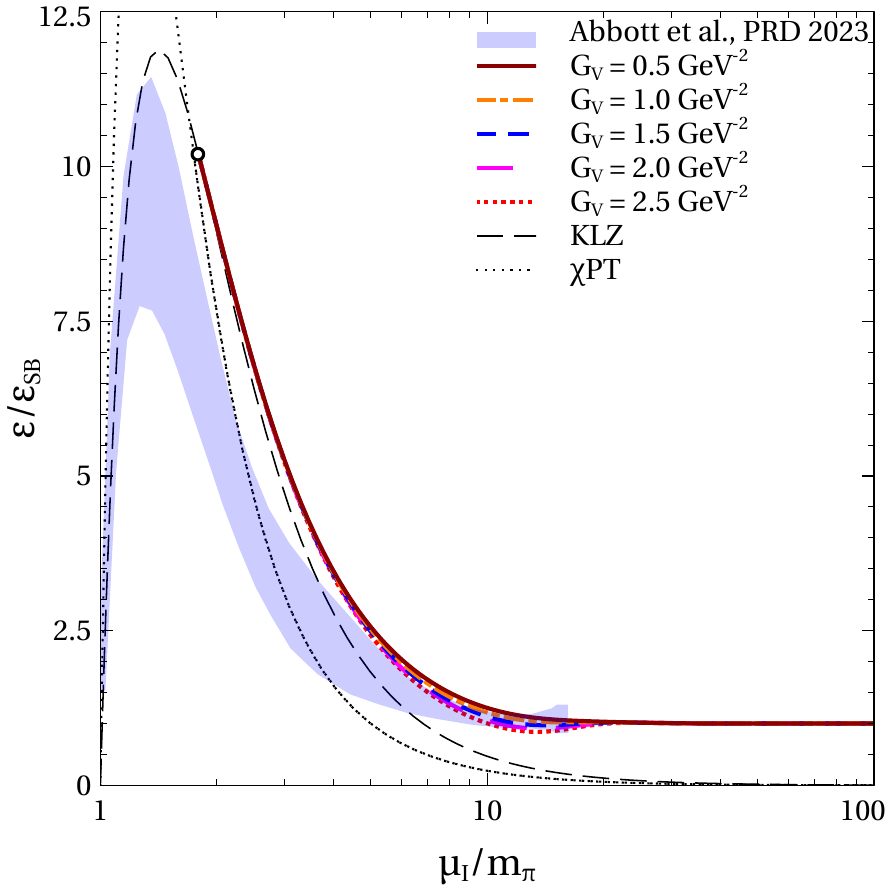}
\caption{Isospin density $n_I$ in units of cube of the vacuum pion mass $m_\pi$ (upper panel) and energy density $\varepsilon$ of cold IQCD matter in units of the energy density of the Stefan-Boltzmann quark gas $\varepsilon_{\rm SB}$ (lower panel) as functions of isospin chemical potential $\mu_I$ in units of the vacuum pion mass.
The calculations are performed within the quarkyonic picture with several values of the vector-isovector coupling $G_V$ indicated in the legend, the KLZ model of interacting pion matter and the EoS obtained within $\chi$PT.
The thin curves on the upper panel represent the contribution of pions.
Empty circles indicate the onset of quarkyonic matter separating these two regimes.
The shaded areas represent the lattice QCD results from Ref. \cite{Abbott:2023coj}.}
\end{figure}

Figure \ref{fig4} demonstrates dependence of the isospin and energy densities of cold IQCD matter on the isospin chemical potential.
Both $n_I$ and $\varepsilon$ exhibit a fast growth right after the onset of the BEC of pions.
At the same time, before the onset of quaryonic matter the KLZ model of pure pion matter overestimates the isospin and energy densities compared to the lattice QCD data from Ref. \cite{Abbott:2023coj}.
This signals that the simple KLZ model of interacting pion matter considered in this work produces a too soft EoS.
This drawback can be improved by accounting for more hadronic states and considering more realistic interactions among them as it is done in the QM model \cite{Chiba:2023ftg,Kojo:2024sca,Andersen:2025ezj,Brandt:2025tkg}. 
Above the onset of quarkyonic matter (anti)quarks gain domination over pions, while the isospin and energy densities coincide with the ones predicted by the lattice QCD \cite{Abbott:2023coj}. 
Thus, the quarkyonic picture of IQCD provides a reasonable agreement with the lattice QCD data.
Is is also seen that above $\mu_{\rm onset}$ pure pion matter described by the KLZ model strongly underestimates isospin and energy densities compared to the quarkyonic picture.

As discussed above, the quarkyonic picture of IQCD predicts asymptotic vanishing of the partial density of pions, which reflects pion dissociation and quark deconfinement in dense medium.
This behavior is seen in the top panel of Fig. \ref{fig4}, which demonstrates that $n_\pi$ decreases after reaching its maximum value already after the onset of quarkyonic matter.
This decreasing and asymptotic vanishing of $n_\pi$ is a striking prediction of the quarkyonic picture of IQCD.
It contrasts with the QM model \cite{Chiba:2023ftg,Kojo:2024sca,Andersen:2025ezj,Brandt:2025tkg}, which treats mesons as fundamental degrees of freedom and, consequently, ignores their dissociation to the composite quarks at any density.
Even dressing the corresponding propagators with quark loops does not provide asymptotic vanishing of the pion condensate in the QM model \cite{Kojo:2024sca}.

\begin{figure}[t]
\label{fig5}
\includegraphics[width=.9\columnwidth]{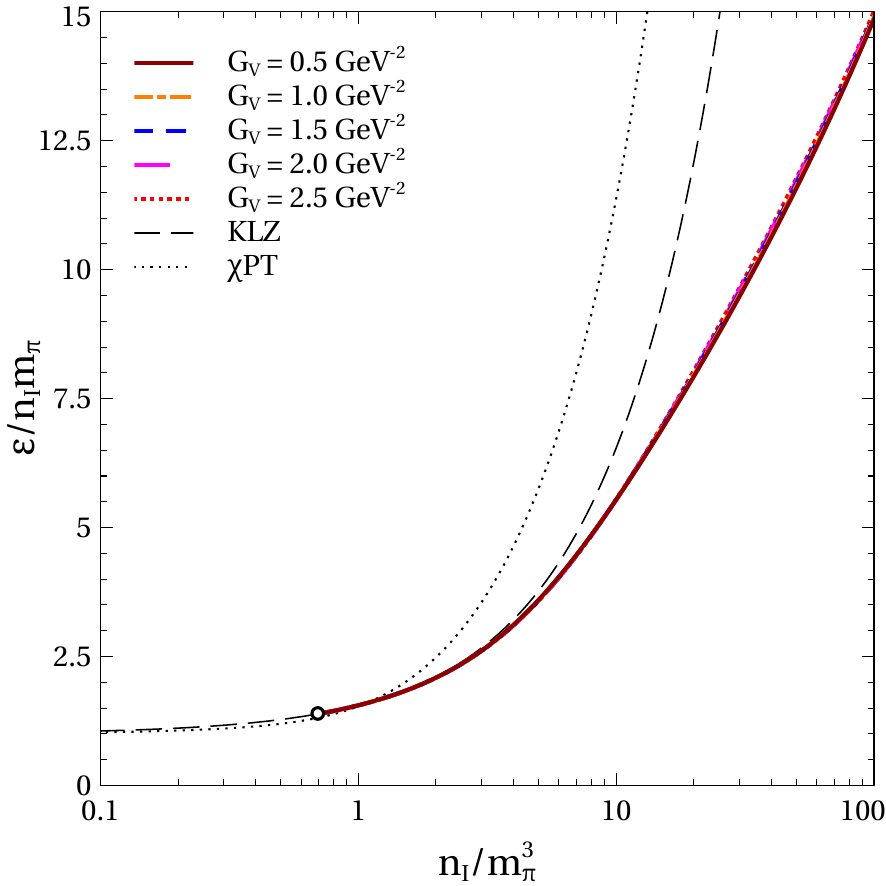}
\caption{Energy per unit of isospin charge $\varepsilon/n_I$ in units of the squared vacuum pion mass $m_\pi$ as a function of isospin chemical potential in units of the vacuum pion mass calculated for the EoSs shown in Fig. \ref{fig4}.}
\end{figure}

It is also interesting to analyze energy per unit of isospin charge $\varepsilon/n_I$ and compare it in the cases of quarkyonic matter and purely pion matter.
In the latter case $k_q=0$ and $\varepsilon/n_I=m_\pi+g^2n_\pi/2m_\rho^2$ at all isospin densities.
Fig. \ref{fig5} shows this quantity as a function of isospin density.
It is seen that above $n_{\rm onset}$ storing isospin charge in the form of quarkyonic matter costs less energy compared to the BEC of pions.
This explains dissociation of pions to their constituent quarks above this density, which is manifested as the onset of quarkyonic matter.
Figure \ref{fig5} also shows that $\varepsilon/n_I$ of quarkyonic matter grows with $G_V$, which means that the vector-isovector repulsion among (anti)quarks increases the energy cost of storing isospin charge in the form of quarkyonic matter.
At the same time, this effect is rather small and does not play a central role in defining the energy budget of cold IQCD matter. 

\begin{figure}[t]
\label{fig6}
\includegraphics[width=.9\columnwidth]{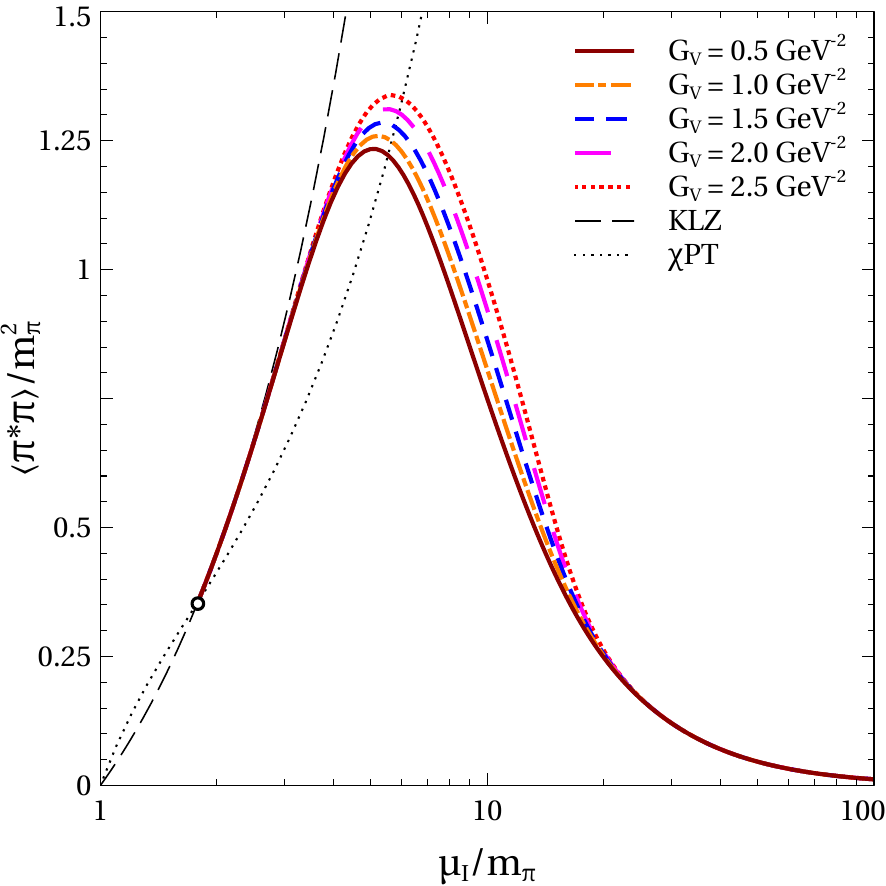}
\caption{Pion condensate $\langle\pi^*\pi\rangle$ of cold IQCD matter in units of the squared vacuum pion mass $m_\pi$ as a function of isospin chemical potential in units of the vacuum pion mass calculated for the EoSs shown in Fig. \ref{fig4}.}
\end{figure}

The present model naturally includes the effects of quark deconfinement and pion dissociation at high densities, which can be seen from the asymptotic vanishing of pion condensate. 
Its dependence on the isospin chemical potential is shown in Fig. \ref{fig6}.
It monotonously grows before the onset of quarkyonic matter and in a limited range of $\mu_I$ after the onset.
Further increase of the isospin chemical potential brings the pion condensate to its maximum with the consequent decrease.
At asymptotically high densities the pion condensate vanishes
$\propto1/\mu_I^2$.
This is in contrast with the QM model predicting $\langle\pi^*\pi\rangle$, which is linear in $\mu_I$ at the tree level and saturates to a finite value at the one-loop level \cite{Chiba:2023ftg,Kojo:2024sca}.
The KLZ model used in this paper also predicts a linear dependence of pion condensate on the isospin chemical potential $\langle\pi^*\pi\rangle=m_\rho^2(\mu_I/m_\pi-1)/2g^2$, which directly follows from Eqs. (\ref{XVI}) and (\ref{XX}) at $k_q=0$.
As mentioned earlier, a pion condensate that does not vanish at asymptotically high densities is a manifestation of ignoring hadron dissociation in the models treating pions as fundamental degree of freedom.
In agreement with the above discussion of the pion shell, Fig. \ref{fig6} also demonstrates that increasing the strength of the vector-isovector repulsion suppresses free (anti)quarks, enhances the fraction of (anti)quarks bound to pions and, consequently, increases the pion condensate.
Thus, determining pion condensate with the first principle lattice QCD simulations opens a possibility of probing repulsion among quarks.

\begin{figure}[t]
\label{fig7}
\includegraphics[width=.9\columnwidth]{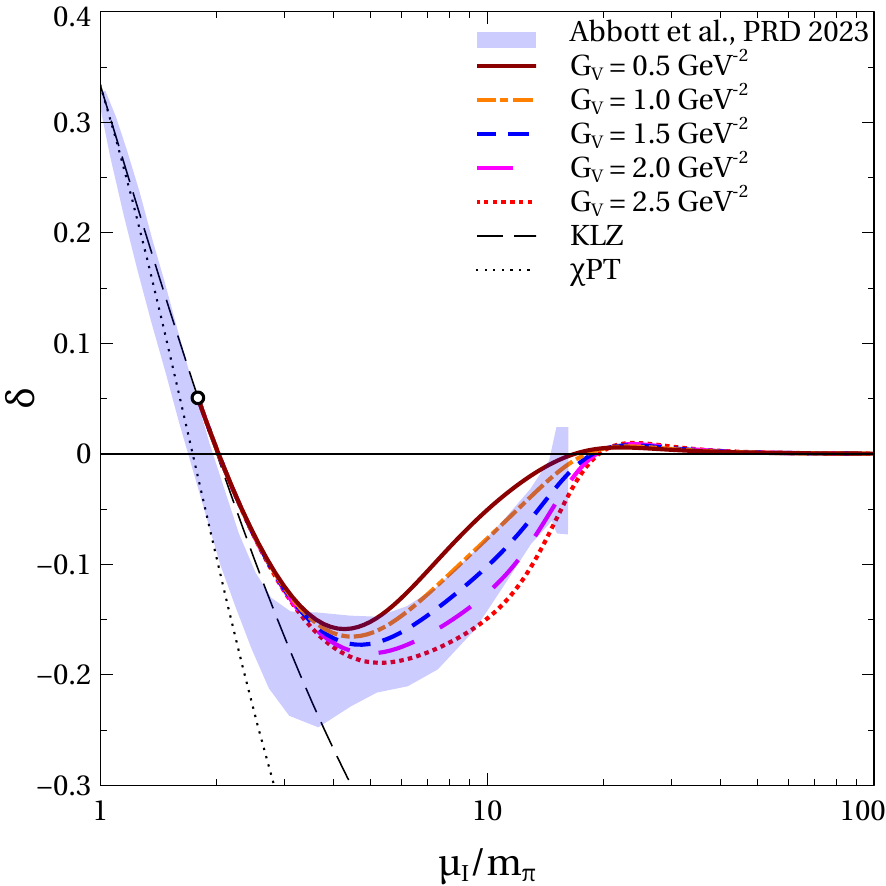}
\includegraphics[width=.9\columnwidth]{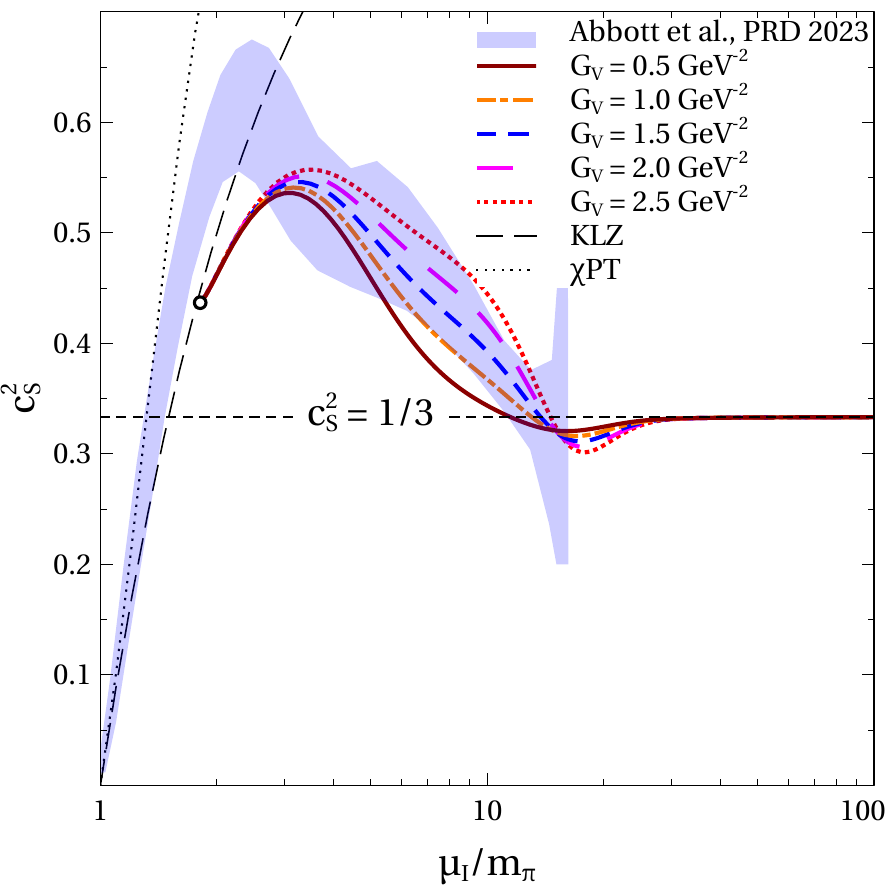}
\caption{Dimensionless interaction measure $\delta$ (upper panel) and squared speed of sound $c_S^2$ (lower panel) of cold IQCD matter as functions of isospin chemical potential $\mu_I$ in units of the vacuum pion mass calculated for the EoSs shown in Fig. \ref{fig4}.
The shaded areas represent the lattice QCD results from Ref. \cite{Abbott:2023coj}.}
\end{figure}

Dimensionless interaction measure and speed of sound characterize stiffness of the EoS of cold IQCD matter and have already been used to analyze its properties \cite{Brandt:2022hwy,Abbott:2023coj,Chiba:2023ftg,Kojo:2024sca,Abbott:2024vhj}.
Fig. (\ref{fig7}) shows these quantities as functions of the isospin chemical potential.
It is rather remarkable that the present simple model provides a reasonable agreement with the lattice QCD data from Ref. \cite{Abbott:2023coj} by adjusting just one freely varying parameter $G_V$.
Before the onset of quarkyonic matter, the KLZ model slightly underestimates stiffness of the BEC of interacting pions, which agrees with the conclusion drawn from the analysis of isospin and energy densities (see Fig. \ref{fig4} and the related discussion).
The stiffness of the IQCD matter keeps growing after the onset of quarkyonic matter.
This is reflected in a moderate decrease of $\delta$ and increase of $c_S^2$ in a range of the isospin chemical potential up to about $3-4$ pion vacuum masses, where these quantities reach their minimum and maximum, respectively.
It was argued in Ref. \cite{Chiba:2023ftg} that saturation of the distribution function of (anti)quarks is responsible for stiffening of the IQCD EoS and developing the peak of its speed of sound.
The saturated low-momentum part of the quark distribution function can be associated with $f_{\bf k}^{\rm free}$ being nonzero in quarkyonic matter only. 
This means that peak of $c_S^2$ is located above the onset of quarkyonic matter, where $f_{\bf k}^{\rm free}\neq0$.
Figure \ref{fig7} agrees with this conclusion.
At the same time, the onset of quarkyonic matter increases $\delta$ and reduces $c_S^2$ of IQCD even in the presence of vector-isovector repulsion among quarks, which signals about its softening compared to the BEC of interacting pions.
At the isospin chemical potentials up to 10 vacuum pion masses the IQCD matter is strongly nonconformal with $\delta<0$ and $c_S^2>1/3$. 
At higher $\mu_I$ about $10-20$ vacuum pion masses it turns to be in a qualitative agreement with perturbative QCD.
At this regime dimensionless interaction measure and speed of sound are above and below their conformal values, respectively.
The conformal limit is reached asymptotically in agreement with Eq. (\ref{XXVII}). 
This is a remarkable outcome of the quarkyonic picture of IQCD since the lattice QCD results do not reach that high chemical potentials yet \cite{Brandt:2022hwy,Abbott:2023coj,Abbott:2024vhj}, while the QM model predicts reaching the conformal limit of $\delta$ and $c_S^2$ from below and above, respectively \cite{Chiba:2023ftg,Kojo:2024sca,Andersen:2025ezj,Brandt:2025tkg}.

%%%%%%%%%%%%%%%%%%%%%%%%%%%%%%%%%%%%%%%%%%%%%%%%%%%%%%%%%%%
\section{Conclusions}
\label{concl}

In this paper we proposed the quarkyonic picture of IQCD inspired by the argument that confining forces among (anti)quarks are important only in the vicinity of their Fermi surface, while deeply in the quark sea these forces are suppressed if the isospin density is high enough. 
Within this picture cold IQCD matter is constituted by $u$ and anti-$d$ quarks and the BEC of the lightest isiospin triplet hadrons, pions.
Heavier hadronic states carrying isospin charge, e.g. charged $\rho$ mesons and nucleons, do not appear since the quarkyonic matter onsets at the isospin chemical potential, which is smaller than the masses of these hadrons.
The corresponding onset density is estimated using the experimentally established charge radius of pions and a simple confining model treating them as the lowest energy $P$-states of three-dimensional harmonic oscillator, where the effects of in-medium screening are accounted for via the geometric overlapping of the potentials.
The remarkable conclusion of the work is that under the conditions of IQCD cold quarkyonic matter can onset at the isospin density about $0.23~\rm fm^{-3}$, which is significantly smaller than the overlap density of pions $0.83~\rm fm^{-3}$. 
The corresponding onset chemical potential is 250 MeV.

Unlike other approaches applied to model cold IQCD matter, the proposed quarkyonic picture directly accounts for quark substructure of pions.
For this the leading order resummed pion propagator is constructed as the four-point correlation function in the random phase approximation.
In agreement with the tenet argument of the quarkyonic picture of IQCD that confining forces among (anti)quarks are important only in the vicinity of the maximum momentum of quarks, only the momentum states from the pion shell of a finite thickness are accounted for to construct the corresponding polarization function.
Evaluating this polarization function at the one-loop no-sea level allows us to analytically find the asymptote of the pole of the pion propagator. 
This asymptote motivates an analytic parametrization of the medium-dependent effective mass of pions, which accounts for their quark substructure. 

To construct EoS of cold IQCD matter we develop a field-theory motivated model of quark-antiquark-pion quarkyonic matter. 
The BEC of pions is described by the KLZ model of pions interacting via the $\rho$-meson exchange, supplemented with the described above medium dependence of the pion mass. 
This medium dependence of the mass of pions couples them to the quark sector, which is modeled as massive fermion gas with nonlocal vector-isovector repulsion.
The latter stiffens the EoS of IQCD  above the onset of quarkyonic matter and provides agreement with the lattice QCD data.
An important result of the work corresponds to determining the thickness of the pion shell in a self-consistent way. 
It is shown that at high densities this thickness scales inverse proportionally to the isospin density. 
This provides asymptotic vanishing of the pion density and pion condensate being a manifestation of pion dissociation, which should be unavoidable due to asymptotic freedom of QCD but is missing in the models treating pions and other hadrons as fundamental degrees of freedom, e.g., linear $\sigma$-model and QM model.

The nonlocal character of the repulsion among (anti)quarks provides reaching the conformal limit of IQCD. 
We demonstrate that in agreement with the predictions of perturbative QCD, speed of sound and dimensionless interaction measure of the developed model approach their conformal values from below and above, respectively.
It is argued that this feature of the quarkyonic picture of IQCD requires the proper scaling of the pion mass, which accounts for their quark substructure, while treating pions as fundamental degrees of freedom leads to contradiction with perturbative QCD.

At the same time, the developed model misses the aspects of dynamical restoration of chiral symmetry and pairing of (anti)quarks, which should be accounted for separately. 
The hadron sector of the model also allows improvements related to accounting for more hadronic species, which can mediate additional interactions between pions and dress their propagators with loop corrections.
Lastly and most important, the findings of the work demonstrate that quark substructure of pions is a crucial element of the IQCD matter.
This motivates development of a microscopic Lagrangian approach to IQCD treating hadrons as (anti)quark correlations, which properties are derived from the underlying quark dynamics.

%%%%%%%%%%%%%%%%%%%%%%%%%%%%%%%%%%%%%%%%%%%%%%%%%%%%%%%%
\vspace{0.5cm}
\section*{ACKNOWLEDGMENTS}
The author acknowledges fruitful discussions with Christoph G\"artlein, Pavlo Panasiuk, Violetta Sagun, David E. Alvarez-Castillo and David Blaschke.
This work was performed within the program Excellence Initiative--Research University of the University of Wrocław of the Ministry of Education and Science and received funding from the Polish National Science Center under grant No. 2021/43/P/ST2/03319. 

%%%%%%%%%%%%%%%%%%%%%%%%%%%%%%%%%%%%%%%%%%%%%%%%%%%%%%%%
\vspace{0.5cm}
\section*{DATA AVAILABILITY}

The data that support the findings of this article are openly available under the condition of referring to it \cite{ivanytskyi_2025_16412194}.

%%%%%%%%%%%%%%%%%%%%%%%%%%%%%%%%%%%%%%%%%%%%%%%%%%%%%%%%%%%
\begin{appendix}
\section{Kroll-Lee-Zumino model}
\label{secApp1}

Within the KLZ model charged pions are described by a complex scalar field $\pi$, which is gauged to the real vector field $\rho_\mu$ representing neutral $\rho$ mesons.
The model Lagrangian is
\begin{eqnarray}
    \label{AI}
    \mathcal{L}=(D_\mu\pi)^*(D^\mu\pi)-M_\pi^2\pi^*\pi-
    \frac{R_{\mu\nu}R^{\mu\nu}}{4}+\frac{m_\rho^2\rho_\mu\rho^\mu}{2},\quad
\end{eqnarray}
where $D_\mu=\partial_\mu-ig\rho_\mu$ and $R_{\mu\nu}=\partial_\mu\rho_\nu-\partial_\nu\rho_\mu$.
Equation (\ref{AI}) leads to the Euler-Lagrange equations
\begin{eqnarray}
    \label{AII}
    &&(D^2+M_\pi^2)\pi=0,\\
    \label{AIII}
    &&\partial_\mu R^{\mu\nu}-m_\rho^2\rho^\nu=gj^\nu.
\end{eqnarray}
Equation (\ref{AIII}) includes the current associated to the conserved charge, which is carried by pions, i.e.,
\begin{eqnarray}
    \label{AIV}
    j^\mu=i\left(\pi^*(D^\mu\pi)-\pi(D^\mu\pi)^*\right).
\end{eqnarray}

We treat the present model within the mean-field approximation. 
This corresponds to replacing the vector field by its expectation value $\langle\rho^\nu\rangle=\rho g^{\nu0}$ with $\rho$ being constant.
We also account that in the BEC phase with nonzero pion condensate $\langle\pi^*\pi\rangle\neq0$ the scalar field is uniform in space. 
This brings it to the form
\begin{eqnarray}
    \label{AV}
    \pi=\sqrt{\langle\pi^*\pi\rangle}e^{i\zeta-ip_0 x^0}
\end{eqnarray}
with $\zeta$ denoting a constant phase, $p_0$ being the single particle energy of pions and $x^0$ standing for the temporal variable.
With this Eqs. (\ref{AII}) and (\ref{AIII}) become
\begin{eqnarray}
    \label{AVI}
    &&p_0=\pm M_\pi-g\rho,\\
    \label{AVII}
    &&\rho=-\frac{gn_\pi}{m_\rho^2}.
\end{eqnarray}
Density of the conserved charge carried by pions in Eq. (\ref{AVII}) is given by the expectation value of $j^0$, i.e.,
\begin{eqnarray}
    \label{AVIII}
    n_\pi=\pm2M_\pi\langle\pi^*\pi\rangle.
\end{eqnarray}
For definiteness below we use the positive solution of $p_0$ with sign ``$+$'', which corresponds to the BEC of particles with positive isospin density and $D_\mu\pi=-iM_\pi\pi g_{\mu0}$.
With this convention and Eq. (\ref{AVIII}) we immediately arrive at the expression for the pion condensate (\ref{XVI}).

Energy density of the system under consideration can be found as the time-time component of the stress-energy tensor.
This gives
\begin{eqnarray}
    \varepsilon_\pi&=&\frac{\partial\mathcal{L}}{\partial(\partial^0\pi)}\partial^0\pi+
    \partial^0\pi^*\frac{\partial\mathcal{L}}{\partial(\partial^0\pi^*)}-\mathcal{L}\nonumber \\
    \label{AIX}
    &=&2M_\pi\langle\pi^*\pi\rangle(M_\pi-g\rho)-\frac{m_\rho^2\rho^2}{2}.
\end{eqnarray}

Using Eqs. (\ref{AVII}) and (\ref{AVIII}) to exclude the pion condensate and vector field, the energy density $\varepsilon_\pi$ can be given the form of Eq. (\ref{XV}).

%%%%%%%%%%%%%%%%%%%%%%%%%%%%%%%%%%%%%%%%%%%%%%%%%%%%%%%%%%%
\section{Quark model with vector-isoscalar repulsion}
\label{secApp2}

Here we present the model of two-flavor quark matter with vector-isovector interaction, which generates a repulsion among quarks at the regimes typical for IQCD.
Similarly to chiral quark models \cite{Scarpettini:2003fj,Blaschke:2007np,Hell:2008cc,Hell:2009by,Hell:2011ic,Ivanytskyi:2024zip}, we model this interaction in a current-current form.
The methods of functional integration provide a powerful and universal tool to solve models of this class in grand canonical ensemble when quark chemical potentials are natural thermodynamic parameters.
However, working in this ensemble is not optimal within the quarkyonic picture, when quark Fermi momenta are the natural thermodynamic parameters. 
Therefore, below we explicitly derive single particle energies, introduce properly normalized wave functions of (anti)quarks and use them to construct the EoS of the present model.

Nonlocality of a repulsive quark interaction is important for reaching the conformal limit of QCD \cite{Ivanytskyi:2024zip}.
Within the separable approximation \cite{Blaschke:1994px,GomezDumm:2001fz} such nonlocality can be absorbed to the space-time dependent form factor $g(z)$, which enters the vector-isovector quark current as
\begin{eqnarray}
    \label{BI}
    \vec{\textfrak{j}}_\mu=\int dz~g(z)\overline q\left(x+\frac{z}{2}\right)\gamma_\mu\vec\tau q\left(x-\frac{z}{2}\right),
\end{eqnarray}
where $q=(u,d)^T$ is two-flavor quark field and $\vec\tau$ is the Pauli matrix vector acting in the space of quark isospin.
The current (\ref{BI}) enters the Lagrangian
\begin{eqnarray}
\label{BII}
\mathcal{L}=\overline{q}(i\slashed\partial-m_q)q-
G_V\vec{\textfrak{j}}_\mu \vec{\textfrak{j}}^\mu.
\end{eqnarray}
We bosonize this Lagrangian by means of the Hubbard-Stratonovich transformation of the partition function $\mathcal{Z}$.
For this we introduce collective vector-isovector field $\vec{\phi}_\mu$ coupled to the current $\vec{\mathfrak{j}}_\mu$.
They appear via the identity
\begin{eqnarray}
\label{BIII}
\mathcal{Z}=\int\mathcal{D}\overline{q}\mathcal{D}q
e^{\int dx~\mathcal{L}}
=\int\mathcal{D}\overline{q}\mathcal{D}q
\mathcal{D}\vec \phi_\mu e^{\int dx~\mathcal{L}^{\rm bos}}
\end{eqnarray}
where the bosonized Lagrangian
\begin{eqnarray}
\label{BIV}
\mathcal{L}^{\rm bos}=\overline{q}(i\slashed\partial-m)q
-\vec{\mathfrak{j}}_\mu\vec{\phi}^\mu+\frac{\vec{\phi}_\mu\vec{\phi}^\mu}{4G_V}.
\end{eqnarray}

At the mean-field level the vector-isovector field is replaced by its expectation value, which is proportional to the expectation value of the vector-isovector quark current.
Only its component generated by the diagonal third Pauli matrix survives the averaging procedure due to the orthogonality of the flavor wave functions of quarks.
With this we obtain $\langle\vec\phi^\mu\rangle=g^{\mu0}(0,0,\phi)$, where $\phi$ is constant.
The corresponding Euler-Lagrange equation includes the expectation value of the third isospin component of the vector-isovector current
\begin{eqnarray}
\label{BV}
\phi=2G_V\langle\mathfrak{j}_{3,0}\rangle.
\end{eqnarray}
Hereafter $\langle\dots\rangle$ means averaging over the spacial three-volume $V\equiv\int d{\bf x}$.

At the next step we switch from the coordinate representation to the momentum one.
This corresponds to expanding the quark fields $q$ in planar waves as
\begin{eqnarray}
    \label{BVI}
    q=\sqrt{V}\int\frac{d{\bf k}}{(2\pi)^3}
    e^{-ikx}\mathfrak{q},
\end{eqnarray}
where $\mathfrak{q}=(\mathfrak{u},\mathfrak{d})^T$ and the factor $\sqrt{V}$ is introduced to normalize them as
\begin{eqnarray}
    \label{BVII}
    \overline{\mathfrak{f}}\gamma_0\mathfrak{f}=
    2N_c\left[\theta(k_f-|{\bf k}|)-
    \theta(k_{\overline{f}}-|{\bf k}|)\right].
\end{eqnarray}
The factor $2N_c$ in Eq. (\ref{BVII}) is the spin-color degeneracy, while $k_f$ and $k_{\overline{f}}$ stand for the Fermi momenta of quarks and antiquarks of the flavor $f$, respectively.

The zeroth component of the four-vector $k_0$ in Eq. (\ref{BVI}) is a three-momentum dependent single particle energy of quark.
To find this $k_0$ we notice that the fields $\mathfrak{q}$ extremize the action, i.e.,
\begin{eqnarray}
   \label{BVIII}
    \frac{\delta}{\delta\overline{\mathfrak{q}}}
    \int dx~\mathcal{L}^{\rm bos}=
    \frac{V}{(2\pi)^2}
    (\slashed{k}-m_q-\tau_3\gamma_0\phi g_{\bf k})\mathfrak{q}=0.
\end{eqnarray}
Up to an insignificant factor this is nothing but the Fourier transformed Dirac equation.
Solving it with respect to $k_0$ we obtain the single particle energies
\begin{eqnarray}
    \label{BIX}
    k^{\pm}_{0f}=\pm(\epsilon_{\bf k}+\phi g_{\bf k}\tau_{3f}),
\end{eqnarray}
where the signs ``$+$'' and ``$-$'' distinguish between quarks and antiquarks, $\tau_{3u}=1$ and $\tau_{3d}=-1$. 

With the above expression for the single particle energies of (anti)quarks we can show that the normalization (\ref{BVII}) provides the proper expression for the net density of the quark flavor $f$
\begin{eqnarray}
    \label{BX}
    \langle\overline{f}\gamma_0f\rangle=
    2N_c\int\frac{d{\bf k}}{(2\pi)^3}
    \left[\theta(k_f-|{\bf k}|)-
    \theta(k_{\overline{f}}-|{\bf k}|)\right].\quad
\end{eqnarray} 
Similarly, we find the net isospin density carried by the quark flavor $f$
\begin{eqnarray}
    n_q=\frac{1}{2}\langle\overline{f}\gamma_0\tau_3f\rangle&=&
    \tau_{3f}N_c\int\frac{d{\bf k}}{(2\pi)^3}\nonumber\\
    \label{BXI}
    &\times&
    \left[\theta(k_f-|{\bf k}|)-
    \theta(k_{\overline{f}}-|{\bf k}|)\right].\quad
\end{eqnarray} 

The energy density of quarks can be found as the average of the time-time component of their energy-momentum tensor, i.e.
\begin{eqnarray}
    \varepsilon_q&=&
    \left\langle\frac{\partial\mathcal{L}}{\partial(\partial^0q)}\partial^0q-\mathcal{L}
    \right\rangle
    \nonumber \\
    &=&2N_c\sum_f\int\frac{d{\bf k}}{(2\pi)^3}
    \left[\theta(k_f-|{\bf k}|)k_{0f}^+-
    \theta(k_{\overline{f}}-|{\bf k}|)k_{0f}^-\right]g_{\bf k}    
    \nonumber \\
    \label{BXII}
    &-&\frac{\phi^2}{4G_V}.
\end{eqnarray}

The vector-isovector field defined by Eq. (\ref{BV}) is obtained similarly
\begin{eqnarray}
    \phi&=&
    4G_VN_c\sum_f\tau_{3f}\int\frac{d{\bf k}}{(2\pi)^3}\nonumber\\
    \label{BXIII}
    &\times&
    \left[\theta(k_f-|{\bf k}|)-
    \theta(k_{\overline{f}}-|{\bf k}|)\right]g_{\bf k}.
\end{eqnarray} 

Noticing that according to the picture described in Sec. \ref{sec2} $k_{u}=k_{\overline{d}}=k_q$ and $k_{\overline{u}}=k_d=0$, we bring Eqs. (\ref{BXI}), (\ref{BXII}) and (\ref{BXIII}) to the form (\ref{XIII}), (\ref{XVII}), and (\ref{XVIII}), respectively.

\end{appendix}

%%%%%%%%%%%%%%%%%%%%%%%%%%%%%%%%%%%%%%%%%%%%%%%%%%%%%%%%%%%
\bibliography{bibliography}
%%%%%%%%%%%%%%%%%%%%%%%%%%%%%%%%%%%%%%%%%%%%%%%%%%%%%%%%%%%

\end{document}